\documentclass[10pt,letterpaper,twocolumn]{article}
\usepackage{ol}
\usepackage{epsfig}
\usepackage{amssymb}
\usepackage{amsmath}
\begin{document}
\twocolumn[ 
\title{Temporal variability of the telluric sodium layer}
\author{D. Saul Davis}
\author{Paul Hickson}
\address{Department of Physics \& Astronomy, University of British Columbia, \\ 6224 Agricultural Road, Vancouver BC V6T 1Z1, Canada}
\author{Glen Herriot}
\affiliation{Herzberg Institute of Astrophysics, \\5071 West Saanich Road, Victoria BC V9E 2E7, Canada}
\author{Chiao-Yao She}
\affiliation{Department of Physics, Colorado State University, \\Fort Collins, CO 80523-1875, U.S.A.}
\begin{abstract}
The temporal variability of the telluric sodium layer is investigated by analyzing 28 nights of data obtained with the Colorado State University LIDAR experiment. The mean height power spectrum of the sodium layer was found to be well fit by a power law over the observed range of frequencies,  $10\ \mu$Hz to $4$ mHz. The best fitting power law was found be be $10^\beta\nu^\alpha$, with $\alpha = -1.79 \pm 0.02$ and $\beta = 1.12 \pm 0.40$.  Applications to wavefront sensing require knowledge of the behavior of the sodium layer at kHz frequencies. Direct measurements at these frequencies do not exist. Extrapolation from low-frequency behavior to high frequencies suggests that this variability may be a significant source of error for laser-guide-star adaptive optics on large-aperture telescopes.
\end{abstract} 
\ocis{350.1260, 110.6770, 010.1330}
 ] 

\section{Introduction}

The advances made in adaptive optics (AO) over the last decade have enabled many ground-based optical astronomical projects that were never before possible\cite{beck}. Most of these studies have been performed with natural guide star AO systems (NGSAO), which require a bright star near to the desired target. While there are many examples of strikingly successful studies done with NGSAO, the fraction of the sky accessible to them is rather small. In order to access the whole sky, one requires the ability to sample the atmospheric distortion anywhere, not just toward the direction of a bright star. The telluric sodium layer is a layer of sodium atoms that is typically present at an altitude of $\sim 90$ km over an extent of $\sim 15$ km. By illuminating the layer with a laser tuned to the wavelength of the $D_2$ transition ($589$ nm), one may excite emission from these sodium atoms. The resulting bright spot acts as an artificial guide star, and enables the operation of laser guide star AO (LGSAO) systems. 

The development of a LGSAO system is of particular importance to the Thirty-Meter telescope (TMT). In order to take full advantage of the potential spatial resolution of the large aperture, it is imperative for many of the TMT's instruments to be fed by the proposed LGSAO system, the Narrow Field InfraRed Adaptive Optic System (NFIRAOS). The development of NFIRAOS presents many technical challenges. As discussed in O'Sullivan et al. (2000)\cite{osull}, one of these challenges---the refocusing of the telescope on the moving laser guide star spot---is due to the variability of the sodium layer. Though the physical mechanism is not entirely understood, atmospheric gravity wave propagating upward in the mesopause region is thought to be the most likely explanation. A typical Shack-Hartmant wavefront sensor will determine the centroid of a illuminated spot on a CCD, and will therefore respond to the instantaneous mean height of the sodium layer. A change in the mean height, $\Delta h$, of the sodium layer produces a RMS wavefront error, averaged over the aperture, 
\begin{equation}
\sigma=\frac{1}{8\sqrt{3}}\left(\frac{D}{h}\right)^2\Delta h,
\end{equation}
where $D$ is the diameter of the telescope aperture, and $h$ is the instantaneous density-weighted mean height of the sodium layer \cite{herr}. Note that $\sigma$ is proportional to the square of the diameter of the telescope. Thus, for the TMT, we expect the effect to be an order of magnitude larger than for the largest optical telescopes presently in operation. For the TMT, $\sigma = 8$ nm for every meter change of the mean height.

While an AO system can use laser and natural guide stars to correct for focus error caused by the atmosphere and variations in the height of the sodium layer, it can not do so instantaneously. Therefore, the correction applied by the AO system will be imperfect, and a residual wavefront error (RWFE) will remain. While the mean height of the sodium layer is known to be variable on the order of several kilometers over timescales of minutes, LGSAO systems sample the atmosphere many times per second. Therefore, it is necessary to know how the mean height of the sodium layer will vary on millisecond timescales in order to estimate the RWFE. In order to assess the variability of the sodium layer on short time scales we examined data taken by the Colorado State University (CSU) Light Detection and Ranging (lidar) experiment over 89 nights in 2003. We determined the typical RWFE expected by multiplying the mean height power spectrum of the sodium layer  by a model rejection transfer function determined specifically for the TMT.

\section{Technique}
\subsection{Reduction and extraction} 
\begin{figure}[!tbp]
\centerline{\epsfig{file=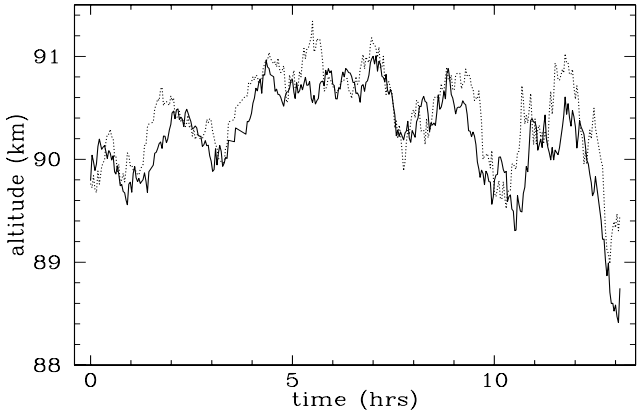}}
\caption{\label{ts.fig} shows the time series of the centroid height for December $24^{th}$. Channel \#1 is shown as a solid line, while channel \#2 is shown as a dotted line.}
\end{figure}
The CSU lidar samples the sodium layer by projecting laser pulses upwards, cyclically tuned to three preselected frequencies within the Doppler-broadened bandwidth of the sodium D$_2$ transition at 589 nm. The intensity of the return signal in three channels (one for each frequency), binned as a function of distance from the telescope can be processed for measuring atmospheric temperatures and horizontal wind between 80 and 105 km\cite{she}. The experiment employs two telescopes that can be independently pointed, and can therefore measure the Na-layer in two directions simultaneously, typically zenith,  due north at an altitude of $60^\circ$, or due east at an altitude of $60^\circ$. In order to increase the signal-to-background ratio, the signal is integrated for a duration of $120$ seconds. Due to readout time, measurements are typically spaced on $125$ second intervals, resulting in a Nyquist frequency of $4$ mHz. 

The signal-to-background ratio was calculated in the following manner: The atmosphere was divided into three regimes. The signal regime extended from $75$ km to $105$ km. Although the sodium layer only extends over approximately half this altitude range at any one time, the large range was necessary to ensure that all the signal was contained in the range regardless of the mean height. The two background regimes extended from $60$ to $72$ km and $108$ to $120$ km respectively. By integrating over the aforementioned regimes a total signal, $S_{total}$, and total background, $B_{total}$, were calculated. The requirements for acceptable data were $S_{total} > 10^4 $ photons and $B_{total} < 10^3$ photons.

 The raw files were processed to yield time, mean height (corrected for zenith angle of the laser, and the altitude of the CSU lidar experiment), and a flag indicating if the signal-to-background ratio of an individual measurement was acceptable. The time series for December $24^{th}$ is shown in figure \ref{ts.fig}. Channel \#1 and channel \#2 are shown as solid and dotted lines respectively.

\subsection{Transformation and Fitting} 
\begin{figure}[!tbp]
\centerline{\epsfig{file=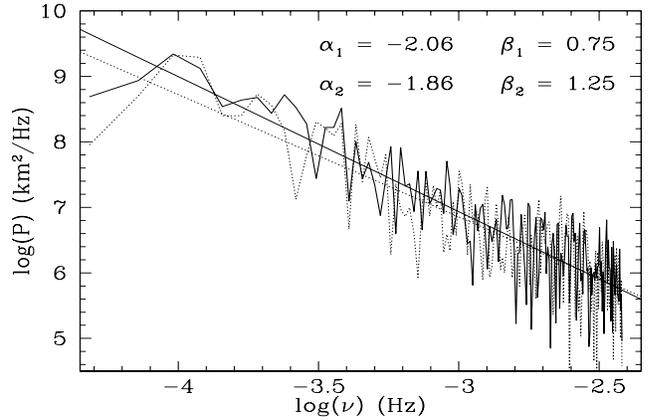}}
\caption{\label{ps.fig}  shows the mean height power spectrum and power-law fits for December $24^{th}$, with $\alpha$ indicating the slope, and $\beta$ indicating the offset.}
\end{figure}
We computed the mean height power spectrum for each night individually. Over the observed range of frequencies, from $\mu$Hz to mHz, the transformed data were well fit by a power law. Many of the nights were split into several runs of acceptable data quality separated by sections of poor-quality data. The extracted power spectrum is a convolution of the Fourier transforms of both the data and the window function. A window function with a broad Fourier transform will yield a power-law fit with an artificially shallow slope, which will in turn predict an inflated value for the RWFE at high frequencies. In order to derive the correct slope of the power-law fit we used only the longest continuous series of measurements from a given night. The power spectrum for the night of December $24^{th}$ is shown in figure \ref{ps.fig}. Channel \#1 and channel \#2 are shown as solid and dotted lines respectively. The slope of the power-law fit for the various channels are $\alpha$ and the offsets are $\beta$. 

\subsection{Determination of residual wavefront error}
\begin{figure}[!tbp]
\centerline{\epsfig{file=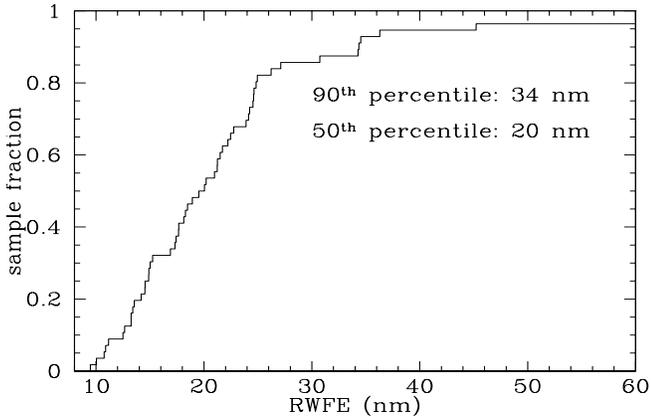}}
\caption{\label{cum7.fig} shows the cumulative distribution for the calculated RWFE for both channels of the 28 nights with more than $7$ hours of data with acceptable data quality.}
\end{figure}
Because the distance to the sodium layer is not know a priori, natural guide stars are required to keep the telescope focused on science targets. In order for a sufficient number of guide stars to be present to make a large fraction of the sky available, the NGSAO system must be designed to work with relatively faint stars. The low fluxes from these stars necessitate a slow read out rate. Therefore, NGSAO can only provide low-frequency information regarding focus changes. A laser guide star can be designed to emit relatively high flux, and can therefore be read out quickly, providing high-frequency focus information.  Herriot et al.\cite{herr} combine the high-frequency focus information from the laser guide star and the low-frequency focus information from the natural guide star to obtain focus information over a wide range of frequencies.  Due to limited bandwidth, the aforementioned correction is not perfect, but may be characterized by the rejection transfer function. The residual wavefront variance (RWFE)$^2$ is proportional to the integral over the appropriate range of frequencies ($10^{-4}$ to $10^4$ Hz) of the product of the mean height power spectrum and the square of the AO system rejection transfer function, 
\begin{equation}
RWFE = \frac{1}{8\sqrt{3}}\left(\frac{D}{h}\right)^2
\left[
\int_{\nu_1}^{\nu_2} h_{Na} \left|  H_{rej}\right|^2df
\right]^{1/2}.
\end{equation}

The residual wavefront error was calculated for each night with a rejection transfer function designed for the TMT. The cumulative distribution for all nights with more than $7$ hours of continuous acceptable data is shown in figure \ref{cum7.fig}. We found that nights with less than $7$ hours of continuous data had a shallower slope than the datasets with longer continuous streams. This is due to the aforementioned effect of the convolution of a narrow window function with the true power spectrum.

\section{Results and Conclusions}
The data for each night with at least one hour of acceptable quality (76 of 89) were analyzed. It was found that nights with sections of usable data that were less than $7$ hours (48 nights) had a shallower slope with respect to those nights with longer sections of usable data (28 nights). Using just the longer nights we find the best-fit power law has the parameters $\alpha=-1.79\pm0.02$ and $\beta=1.12\pm0.40$. We found that RWFE for those nights has a median value of $20$ nm and a $90^{th}$ percentile value of $34$ nm. This effect is an important contribution to the overall error budget of NFIRAOS, and more generally, will likely be for most, if not all, of the LGSAO systems designed for the next generation of large ground-based telescopes. 

This analysis was limited by the relatively long integration times of the CSU lidar experiment, which in turn set a low Nyquist frequency. The variability of the sodium layer height in the frequency range observed is mainly due to the upward passing of atmospheric waves. Atmospheric turbulence at the sodium layer height can perturb the mean height of the sodium layer on time scales of seconds or shorter. However, changes caused by atmospheric turbulence are likely to be small (cm or less), unlike the kilometer-scale changes reported here. The assumption that the mean height power spectrum can be fit by a single unbroken power law from $\mu$Hz to kHz is perhaps na\"ive, but reasonable. The actual high-frequency behavior of the mean sodium height power spectrum, of course, can only be determined by further experimentation. The temporal resolution could be improved by designing a lidar experiment with either a larger collecting telescope, or by sacrificing spatial resolution. 

\section*{Acknowledgments}
This research was supported by the Natural Sciences and Engineering Research Council of Canada. The work at Colorado State University was in part supported by National Aeronautics and Space Administration, under grant NAG5-10076 and National Science Foundation, under grant ATM-00-03171. \@ S. Davis' email is {\tt sdavis@astro.ubc.ca}.

\end{document}